\documentclass[prl,twocolumn,showpacs,preprintnumbers,amsmath,amssymb]{revtex4}

\usepackage{graphicx}
\usepackage{subfigure} 
\usepackage{dcolumn}
\usepackage{bm}

\begin{document}

\draft

\title{Density dependent exchange contribution to
  $\partial\mu/\partial n$ and compressibility in graphene} 
\author{E.~H.~Hwang,$^1$ Ben Yu-Kuang Hu,$^{2,1}$ and S.~Das Sarma$^1$}
\address{$^1$Condensed Matter Theory Center, 
Department of Physics, University of Maryland, College Park,
Maryland  20742-4111} 
\address{$^2$Department of Physics,
University of Akron, Akron, OH 44325-4001}
\date{\today}

\begin{abstract}
We calculate $\partial\mu/\partial n$ (where $\mu = $ chemical
potential and $n =$ electron density), which is associated with the
compressibility,  
in graphene as a function of $n$, within the
Hartree-Fock approximation.   
The exchange-driven Dirac-point logarithmic singularity in the
quasiparticle velocity of intrinsic graphene disappears in the
extrinsic case.  
The calculated renormalized $\partial\mu/\partial n$ in extrinsic
graphene on SiO$_2$ has the same $n^{-\frac12}$ 
density dependence but is 20\% larger than the inverse bare density of
states, a relatively weak effect compared to the corresponding
parabolic-band case.   
We predict that the renormalization effect can be enhanced to about
50\% by changing the graphene substrate. 
\pacs{81.05.Uw, 71.70.Gm; 71.10.-w; 71.18.+y}

\end{abstract}
\maketitle

The band structure of graphene (a single layer of carbon atoms), by
dint of its honeycomb lattice,  
has linear dispersions near the ${\cal K}$ and ${\cal K}'$ points
(``Dirac points") of the Brillouin zone. 
Recent developments in techniques for fabricating conducting graphene
layers have thus provided the physics community with a unique
opportunity  
to study an interacting two-dimensional (2D) massless Dirac fermion
system using table-top experimental equipment. 
This has led to a veritable explosion of both experimental and
theoretical activity in this field\cite{Geim07}. 

Around the Dirac points (which we take to be the zero of energy), the
kinetic energy for a ``bare" electron (see below) is  
$\epsilon^{(0)}_{\bm k,s} = s v_0 |\bm k|$, 
where $\bm k$ is the wavevector with respect to the Dirac point, and
$s = +1$ and $-1$ for the conduction and valence bands, 
respectively.  The electron chemical potential $\mu$, which in
intrinsic graphene is at zero,  
can be shifted up or down by doping and/or application of external
gate voltages, with a concommitant change in the electron density.  
This paper reports the calculation of $\partial\mu/\partial n$, which
is related to the electronic compressibility,  
in extrinsic graphene at temperature $T=0$ as a function of the density $n$. 
[In this paper, unless otherwise indicated, partial derivatives are at
constant area and $T=0$, 
and $n$ refers exclusively to the {\em free} carrier density ({\em
  i.e.}, the difference in electron density from that of intrinsic
graphene)  
in the gated graphene, which we take to be substantially less ($|n|
\sim 10^{12}\,{\rm cm}^{-2}$)  
than the intrinsic electron density $n_v$ ($> 10^{15}\,{\rm cm}^{-2}$)
filling up the valence band.] 
We obtain $\mu(n)$ by evaluating the electron self-energy within the
Hartree-Fock approximation (HFA). 
The HFA is a good approximation up to reasonably high values of $r_s$
($\sim$ the ratio of the average carrier potential to kinetic energy)  
in parabolic-band semiconductors, and we expect it to also give
reliable results in graphene, where $r_s < 1$. 

It is useful (and conceptually meaningful) to divide 2D graphene into
three different systems depending on the band filling: {\em bare} or
{\em empty},  
a theoretical abstraction of just one electron in the graphene
honeycomb lattice as appropriate for the single-particle
band-structure calculation  
with both valence and conduction bands completely (and unphysically)
empty, or equivalently, the unphysical situation where the interaction
between 
the electrons is turned off; {\em intrinsic}, {\it i.e.}, the undoped
and ungated situation,  
which is a zero-gap semiconductor with a completely full (empty)
valence (conduction) band and chemical potential $\mu$ ($= E_F$, since
we are at $T=0$)  
precisely at the Dirac point; {\em extrinsic}, {\it i.e.}, gated/doped
graphene with a tunable 2D  
free carrier density $n$ of electrons (holes) in the conduction (valence)
band, with $\mu$ being above (below) zero, {\em i.e.} in the
conduction (valence) band.   
Note that only the empty system can be characterized by the bare,
noninteracting parameters ({\em e.g.}, velocity $v_0$, density of
states $D_0$) with 
both intrinsic and extrinsic graphene being characterized by
renormalized parameters.  We emphasize that the bare graphene
parameters, being 
unphysical abstractions, cannot be {\em experimentally} determined.

In the absence of interaction, $\partial\mu/\partial n$  
is just the inverse of the bare or non-interacting single-particle
density of states at the 
Fermi level: $(\partial n/\partial\mu)_{0} \equiv D_0(E_F) = \sqrt{g
  n}/(\sqrt{\pi} \hbar v_0)$, where $v_0$ is the bare graphene carrier  
velocity associated with the linear energy dispersion, and $g (= 4)$
is the product of the spin ($g_{s} = 2$) and valley ($g_{v} = 2$)
degeneracies of the graphene carriers. 
This implies that $\partial\mu/\partial n$ in graphene is a direct
measurement of the {\em thermodynamic} Fermi velocity renormalization
due to electron--electron 
interaction effects. (This should be distinguished from the {\em
  quasiparticle} Fermi velocity renormalization, as discussed later.)  
Our goal here is to theoretically calculate the renormalized
$\partial\mu/\partial n$ in intrinsic and extrinsic graphene including
exchange interaction 
effects, or equivalently in the HFA, which should be an excellent
quantitative approximation in 2D graphene. 
Our calculated carrier density dependence of $\partial\mu/\partial n$
can be directly compared to experimental measurements in extrinsic
graphene.

The exchange self energy is given by\cite{Mahan}
\begin{equation}
\Sigma_{{\rm x},s}(k) = -\sum_{s' {\bm q}} 
n_F(\xi_{\bm k-\bm q,s'})\, V_c(\bm q)\, F_{ss'}(\bm
k, \bm k-\bm q), 
\label{eq:1}
\end{equation}
where  $s,s' = \pm 1$ are the band indicies, and since we assume
$T=0$, the fermi function $n_F(\xi) = 0$ or $1$ for $\xi \equiv
\epsilon - \mu$ less than or greater than 0, respectively.  
$V_c(q) = 2\pi e^2/(\kappa q)$ is the bare coulomb potential  
($\kappa$ is the background dielectric constant in the graphene layer).  
$F_{ss'}(\bm k,\bm k')=(1 + ss'\cos\theta_{\bm k\bm k'})/2$ arises
from the wavefunction overlap factor, 
where $\theta_{\bm k\bm k'}$ is the angle between $\bm k$ and  
$\bm k'$. We assume that the valence band is cut off at the wavevector
$k_c$ with respect to the Dirac point.
The ultra-violet regularization associated with the wavevector cut-off
at $k_c$ happens at a very large wavevector, comparable to the lattice
wavevector; 
$k_c \sim 1{\rm \AA}^{-1}$.
Within the HFA, $\mu = \epsilon^{(0)}_{k_F,s} + \Sigma_{{\rm
    x},s}(k_F)$, where $k_F = (4\pi\,|n|/g)^{\frac12}$ is the Fermi
wavevector.   

We separate the exchange self-energy into contributions from the
intrinsic electrons, 
$\Sigma_{\rm x}^{\rm int}$, 
and the extrinsic carriers, $\Sigma_{\rm x}^{\rm ext}$. That is,
$\Sigma_{{\rm x},s}(k) = \Sigma_{{\rm x},s}^{\rm int}(k) +  
\Sigma_{{\rm x},s}^{\rm ext}(k)$, where
\begin{subequations}
\begin{eqnarray}
\Sigma_{{\rm x},s}^{\rm int}(k) & = & -\sum_{{\bm q}}
V_c(\bm q)\, F_{s,-}(\bm k, \bm k-\bm q); \\ 
\Sigma_{{\rm x},s}^{\rm ext}(k) & = & -\sum_{s'{\bm q}}
\delta n_F(\xi_{\bm k - \bm q,s'})
V_c(\bm q)\,F_{ss'}(\bm k, \bm
k-\bm q)
\end{eqnarray}
\end{subequations}
where $\delta n_F(\xi_{\bm k - \bm q,s'}) = n_F(\xi_{\bm k- \bm q,s'})
- \frac{1}{2}(1 - s')$ is the difference in the electron occupation  
from the intrinsic $T=0$ case.
Evaluating the integrals, we obtain 
\begin{subequations}
\begin{eqnarray}
\Sigma_{{\rm x},s}^{\rm int}(k) &=& \frac{e^2 k_c}{\pi\kappa}
\left[-f\!\left(\frac{k}{k_c}\right) +
  s\,h\!\left(\frac{k}{k_c}\right)\right],\label{eq:3a}\\ 
\,\Sigma_{{\rm x},s}^{\rm ext}(k) &=&  \frac{e^2
  k_F}{\pi\kappa}\left[\mp f\!\left(\frac{k}{k_F}\right) - s\,
  h\!\left(\frac{k}{k_F}\right)\right], 
\label{eq:3b}
\end{eqnarray}
\end{subequations}
where $\mp$ in Eq.~(\ref{eq:3b}) is for $\mu \gtrless 0$, 
\begin{equation}
f(x) = 
\begin{cases} 
E(x) & \text{if $x\le 1$;} \\
xE(\frac{1}{x}) - \left(x-\frac{1}x\right) K\left(\frac{1}x\right)
&\text{if $x> 1$,} 
\end{cases}
\end{equation} 
and
\begin{equation}
h(x) = 
\begin{cases}
x \left [\frac{\pi}{4}\log(\frac{4}{x}) - \frac{\pi}{8} \right ] -
x\int_0^x dy\ y^{-3}  \\ 
\;\;\;\;\;\; \times [K(y) - E(y) - \frac{\pi}{4} y^2], &
\text{for $x \le 1$}; \\
x \int_0^{x^{-1}} dy\ [K(y) - E(y)], & \text{for $x > 1$}.
\end{cases}
\label{eq:5}
\end{equation}
Here, $K(x)$ and $E(x)$ are the complete elliptic integral of the
first and second kinds, respectively\cite{Gradshteyn}. 
Note that the $T=0$ exchange self-energy for a regular parabolic-band
two-dimensional electron gas (2DEG) is \cite{Chaplik71} 
$\Sigma_{{\rm x}}^{\rm pb}(k) = -\frac{2e^2
  k_F}{\pi\kappa}\,f\!\left(\frac{k}{k_F}\right)$.
At $T=0$, $\Sigma_{\rm x}(k)$ does not depend on the band-structure
away from the Fermi surface (since $n_F$ is either $1$ or $0$ for 
$\xi < 0$ and $\xi > 0$ respectively, independent of the details of
the band-structure) and therefore the {\sl only} difference between  
$\Sigma_{\rm x}(k)$ for the parabolic-band case and and the intraband
contribution for graphene is the difference in the wavefunction
overlap factor  
$F_{ss'}({\bm k,\bm k'})$.  This accounts for presence of the
$f(k/k_F)$ in both the $T=0$  
expressions for $\Sigma_{\rm x}^{\rm pb}(k)$ and $\Sigma_{\rm x}^{\rm ext}(k)$.
 
First, we examine the intrinsic self-energy, $\Sigma^{\rm int}_{\rm x}(k)$,  
which is independent of the carrier density $n$. 
Since we are interested in the states around the Dirac point, the argument
of the functions $f$ and $h$ in Eq.~(\ref{eq:3a}),  $k/k_c \ll 1$.
For small $x$, $f(x) = E(x) \approx \frac{\pi}2 - O(x^2)$, and $h(x) \approx 
x(\frac{\pi}{4}\log(\frac{4}{x}) - \frac{\pi}{8}) + O(x^3)$
(in Eq.~(\ref{eq:5}), the integrand $y^{-3}[K(y) - E(y) -
\frac{\pi}{4}y^2] \sim y$ as $y\rightarrow 0$, and therefore the
integral 
$\sim x^2$ for small $x$).  Therefore,
\begin{equation}
\Sigma_{{\rm x},s}^{\rm int}(k) = \frac{e^2}{\kappa}
\left\{-\frac{k_c}{2} + s \frac{k}{4}
  \left[\log\left(\frac{4k_c}{k}\right) - \frac{1}{2} +
    O\left(\frac{k}{k_c}\right)\right]\right\} 
\label{eq:7}
\end{equation} 
The term $-e^2 k_c/(2\kappa)$ in Eq.~(\ref{eq:7}) simply shifts energy
zero and can be ignored. 
The other terms renormalize the quasiparticle velocity.  Ignoring
terms of order $k/k_c$  
the renormalized quasiparticle velocity is \cite{Guinea_DasSarma,Barlas}
\begin{equation}
v_{\rm int}(k) = \frac{\partial[\epsilon_{k,s}^{(0)} + \Sigma_{{\rm
      x},s}^{\rm int}(k)]}{\hbar\, \partial k}  
= v_0\left[1 +
  \frac{r_s^{(0)}}{4}\log\left(\frac{\tilde{k}_c}{k}\right)\right], 
\label{eq:8}
\end{equation}
where $\tilde{k}_c \equiv 4e^{-\frac32} k_c \approx 0.9 k_c$ and
$r_s^{(0)} = e^2/(\hbar\kappa v_0)$. 
Experimental measurements of the quasiparticle velocity in intrinsic
graphene will yield $v_{\rm int}$ (in the absence of phonon coupling),  
and not the bare velocity $v_0$, which applies only for the
unrealistic situation of a completely empty valence-band. 
This situation is analogous to the quantum electrodynamics calculation
of the self-energy of a bare electron. The  
bare electron charge and mass of the theory are never observed.  
Instead, experimentally one sees the scale-dependent renormalized
charge and mass, which include effects of the electron self-energy. 
The logarithmic dependence of the intrinsic graphene velocity is
probably difficult to observe because of the smallness of the
prefactor  
$r_s^{(0)}/4 \approx 0.2$ for graphene mounted on a SiO$_2$ substrate
with one side exposed to air  
(hence, the effective $\kappa$ in the graphene layer is the average of
the $\kappa$ of air and SiO$_2$, $\approx 2.5$).     To see   
clearly the logarithmic dependence in Eq.~(\ref{eq:8}), $k$ must be
varied over a fairly wide range.   Furthermore, the  
logarithmic divergence in $v_{\rm int}(k)$ at $k\rightarrow 0$ occurs
{\em only} in the intrinsic graphene, and {\em not} in the extrinsic
case. 

In extrinsic graphene, $k_F \ne 0$.  For $k/k_F \ll 1$, the small $x$
expansions for $f(x)$ and $h(x)$ in Eq.~(\ref{eq:3b}) yield 
\begin{equation} 
\Sigma_{{\rm x},s}^{\rm ext}(k) = \frac{e^2}{\kappa}\left\{
\mp \frac{k_F}{2} - s \frac{k}{4}
\left[\log\left(\frac{4k_F}{k}\right) - \frac{1}{2} +
  O\left(\frac{k}{k_F}\right)\right]\right\}, 
\end{equation}
(where $\mp$ is for $\mu \gtrless 0$). The $\log(k)$ term in 
$\Sigma_{{\rm x},s}^{\rm ext}(k\rightarrow 0)$ cancels
the equivalent term in $\Sigma_{{\rm x},s}^{\rm int}(k\rightarrow 0)$,
so the sum, $\Sigma_{{\rm x},s}(k) = \Sigma_{{\rm x},s}^{\rm ext}(k) +
\Sigma_{{\rm x},s}^{\rm int}(k)$,
has a finite derivative at $k=0$, and the renormalized velocity in the
extrinsic case, 
$v_{\rm ext}(k=0) = v_0 \left[1  + \frac{r_s^{(0)}}{4}
  \log\left(\frac{k_c}{k_F}\right)\right],$ 
has no $k\rightarrow 0$ logarithmic divergence.

Fig.~1 shows $\Sigma_{\rm x}^{\rm ext}(k)$ for graphene with $\mu > 0$
and, for comparison, $\Sigma_{\rm x}^{\rm pb}(k)$. 
At $k=k_F$ (and $\mu>0$),
\begin{equation}
\Sigma_{{\rm x},s}^{\rm ext}(k_F) =
\frac12\left[1+s\left(C-\frac12\right)\right]\Sigma_{\rm x}^{\rm
  pb}(k_F), 
\label{eq:12}
\end{equation}
where $C \approx 0.916$ is Catalan's constant, and $\Sigma^{\rm
  pb}_{\rm x}(k_F) = -\pi e^2 k_F/(\pi\kappa)$.   
As in the case of $\Sigma_{\rm x}^{\rm pb}(k)$, the slope of
$\Sigma_{\rm x+}^{\rm ext}(k)$  
for graphene with $\mu > 0$ has a logarithmic divergence as $k
\rightarrow k_F$ from both the $f(x)$ and $h(x)$ terms in
Eq.~(\ref{eq:3b}).   
We expect that
the logarithmic divergence in $d\Sigma^{\rm ext}_{{\rm x}+}/dk$ will
disappear when correlation effects are included, as in the case 
of the parabolic-band $\Sigma^{\rm pb}_{\rm x}$. 
Note that this logarithmic divergence has no singular pathological
effect on $\partial\mu/\partial n$, the quantity of interest in this
work, and is irrelevant for our purpose.
The $\Sigma_{\rm x-}^{\rm ext}$ has a finite derivative at $k=k_F$,
because for an electron-doped sample there is no Fermi surface at
$k=k_F$ in the 
valence band.

\begin{figure}
\includegraphics[scale=0.35]{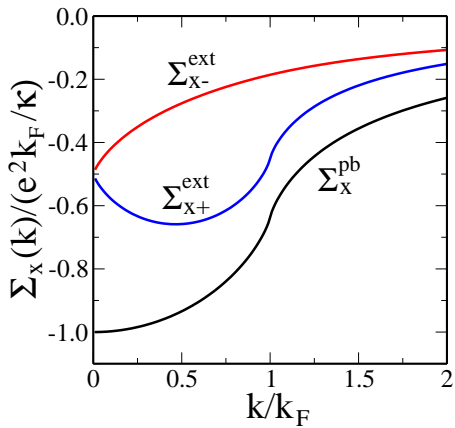}
\caption{(Color online)
Exchange self-energies for graphene (with $\mu > 0$), and for a
parabolic-band 2DEG, as functions of wave vector.   
Note that $\Sigma_{{\rm x}+}^{\rm ext}(k) + 
\Sigma_{{\rm x}-}^{\rm ext}(k) = \Sigma_{\rm x}^{\rm pb}(k)$.
}
\end{figure}

We now discuss the effect of the exchange self-energy on
$\partial\mu/\partial n$.  
In a regular parabolic-band 2DEG with
mass $m$, $\left(\frac{\partial \mu}{\partial n}\right)^{\rm pb}_{0} =
\frac{2\hbar^2 \pi}{m g_s g_v}$ is density independent. When
interactions are included 
this picture changes due to exchange and correlation effects of the
Coulomb potential. Within the HFA, which neglects correlation effects, 
$\left(\frac{\partial \mu}{\partial n}\right)^{\!\!\rm pb} =
\left(\frac{\partial \mu}{\partial n}\right)_{\!\!0}^{\rm\!\!pb}  
\left ( 1 -
  \frac{\sqrt{2}}{\pi} r_s^{\rm pb} \right )$,
where $r_s^{\rm pb} = \sqrt{2} e^2 m/(\kappa k_F) \propto n^{-\frac12}$.
Thus, $\partial \mu/\partial n$ becomes negative at low enough
densities. Measurements of $\partial \mu/\partial n$ in two-dimensional 
electron and hole gases have confirmed this
behavior\cite{PhysRevLett.68.674,PhysRevLett.77.3181}. 
The observed change of sign in $\partial\mu/\partial n$ comes mainly
from the exchange 
contribution to the total energy. It is known that
the correlation corrections to $\partial \mu/\partial n$ beyond the HFA
is not very large ($< 20 \%$) \cite{PhysRevB.39.5005}, even at the
reasonably large effective $r_s$ ($\gg 1$) values at which the 2D
semiconductor experiments 
have typically been carried out.

What is the contribution of exchange on $\partial\mu/\partial n$ in graphene?  
Using $k_F = \sqrt{\pi\,|n|}$, together with $\mu_s(k_F) =
\epsilon^{(0)}_{k_F,s} + \Sigma_{{\rm x},s}^{\rm int}(k_F) +
\Sigma_{{\rm x},s}^{\rm ext}(k_F)$, 
and Eq.~(\ref{eq:12}) gives 
\begin{eqnarray}
\left(\frac{\partial \mu}{\partial n}\right)_{\!\!\rm ext} 
&=& \frac{\sqrt{\pi}}{2\sqrt{|n|}}\left\{\hbar v_0 +
  \frac{e^2}{\kappa}\left[\frac14\log\left(\frac{\tilde{k}_c}{k_F}\right) -  
\frac{C+\frac12}{\pi}\right]\right\}\nonumber\\
&=& \left(\frac{\partial \mu}{\partial n}\right)_{\!\!\rm int}\left[1
  - \frac{C+\frac{1}{2}}{\pi} r_s^{\rm int}\right], 
\label{eq:14}
\end{eqnarray}
where $r_s^{\rm int} = e^2/(\kappa \hbar v^{\rm int}_{F})$  [here,
$v^{\rm int}_{F} \equiv v_{\rm int}(k_{F})$] and 
$(\partial \mu/\partial n)_{\rm int} = (\hbar v^{\rm int}_{F}
\sqrt{\pi})/(2\sqrt{|n|})$ 
is the inverse of the density of states for intrinsic graphene.
This shows that in extrinsic graphene, the exchange effect changes
$\partial\mu/\partial n$ from the 
bare value $(\partial\mu/\partial n)_{0}$ by a factor of
$r_s^{(0)}\left[\frac14\log\left(\frac{\tilde{k}_c}{k_F}\right) -  
\pi^{-1}(C+\frac12)\right]$, or from the intrinsic value
$(\partial\mu/\partial n)_{\rm int}$ by a factor of  
$-\pi^{-1}(C+\frac{1}{2})r_s^{\rm int}$.
Thus, $(\partial\mu/\partial n)_{\rm ext}$ is {\em enhanced} over 
$(\partial\mu/\partial n)_{0}$ by $\approx (0.25)\,r_s^{(0)}$ (for
$n=10^{12}\,{\rm cm}^{-2}$),  
but {\em reduced} from $(\partial\mu/\partial n)_{\rm int}$ by a
factor of $\approx (-0.45)\,r_s^{\rm int}$. 
If we take single-particle band-structure graphene velocity
$v_0\approx 10^{8}\,{\rm cm/s}$, we get $r_s^{(0)} \approx 0.9$  
for SiO$_2$ mounted graphene, giving an exchange enhancement of
approximately 20\% over the bare $(\partial\mu/\partial n)_0$; see
Fig.~2(a).   
Estimating the change with respect to $(\partial \mu/\partial n)_{\rm int}$ 
is a little trickier because $r_s^{\rm int}$, which depends on the
intrinsic graphene velocity $v_F^{\rm int}$, is  unknown since the
intrinsic  
graphene velocity is at present unknown!
An approximate way to estimate the instrinsic $r_s^{\rm int}$ is to
change $\kappa \rightarrow 
\kappa\kappa^*$ where $\kappa^*$ is the effect of the background
screening by the filled valence band\cite{condmat0610561},  
with $\kappa^* = 1 + \frac\pi{8} g_{\rm s}g_{v} r_s^{(0)} \approx 2$,
which gives $r_s^{\rm int} \approx r_s^{(0)}/2$,  
leading to around a 20\% decrease of $(\partial\mu/\partial n)_{\rm
  ext}$ with respect to $(\partial \mu/ 
\partial n)_{\rm int}$.  The extrinsic $\partial\mu/\partial n$
depends on $\kappa$, as shown  
in Fig.~2(b), but unlike the parabolic band case, for experimentally
relevant parameters $(\partial\mu/\partial n)_{\rm ext}$ does not
change sign. 
By using freely suspended graphene ({\em i.e.}, $\kappa = 1$), the
many body renormalization can be enhanced to around 50\%.

\begin{figure}
\includegraphics[scale=0.3]{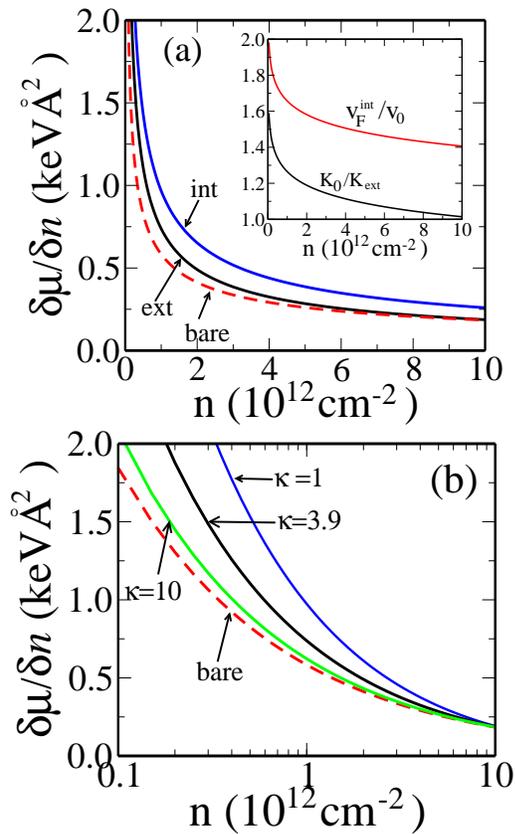}
\\
\includegraphics[scale=0.4]{fig_2b.eps}
\caption{ (Color online)
(a) Calculated $\partial\mu/\partial n$ as a function of free carrier
density, using the 
following parameters: $k_c = 1/a$ ($a=$2.46 \AA), $\kappa = 2.5$, $v_0
= 10^8\,{\rm cm/s}$, and $r_s^{(0)} = 0.9$.  
The ``bare" curve is $\partial\mu/\partial n$ of a noninteracting
graphene, 
and the ``int" and ``ext" curves are   
for the intrinsic and extrinsic cases, respectively.
The inset shows the ratios of the renormalized intrinsic velocity at
$k_F$ and the inverse of the extrinsic compressibility to their
corresponding bare quantities. 
(b) $(\partial\mu/\partial n)_{\rm ext}$ for different values of
$\kappa$ (hence, different $r_s$).  
}
\end{figure}

In the literature, $\partial\mu/\partial n$ is often associated with the term ``compressibility," defined as 
$K \equiv -V^{-1}(\partial V/\partial P)_{T,N}$,  
where $N$ is particle number, $V$ is the system volume/area, $P\equiv -(\partial {\cal F}/\partial V)_{T,N}$ is the pressure, and 
${\cal F}$ is the Helmholtz free energy.  It can be shown that\cite{Schwabl} $K^{-1} = n^2 (\partial \mu/ \partial n)$, where $n = N/V$. 
In experimental papers on the compressibility of electron gases, the quantity that is
measured is not actually the compressibility (after all,
experimentalists do not  
physically compress the electron gas and measure the change in pressure) 
but $\partial\mu/\partial n$, which is then converted to $K^{-1}$ by multiplication of $n^2$.  
In graphene, it is in fact ambiguous which $n$ should be used --- (a) the free carrier density or (b) the density of the electrons in the band (i.e.,
free carrier density plus $n_v$)?  The answer depends on which hypothetical compressibility is being considered --- (a) corresponds to one 
in which the area enclosing the free carriers is changed but the underlying lattice is kept constant, and (b) to one where the volume of the 
underlying lattice (and hence $n_v$) also changes.  To avoid any ambiguities, we use the quantity $\partial\mu/\partial n$. 

Before concluding, we point out that it is incorrect to think of $(\partial\mu/\partial n)_{\rm ext}$
as providing a measurement of many-body quasiparticle Fermi velocity renormalization in graphene.  
In particular, the quasiparticle velocity renormalization is given by the $F_1^{s}$ parameter in Fermi liquid theory through Galelian invariance
$v_F/v_F^{\rm ren} = 1 + F_1^s$, whereas the renormalization of $\partial\mu/\partial n$ 
is related to the Fermi liquid parameter $F_0^s$ through the identity
\begin{equation}
\left(\frac{\partial\mu}{\partial n}\right)^{\!\!\rm ren} = \left(\frac{\partial\mu}{\partial n}\right)\,(1 + F_0^s) \frac{v_{F}^{\rm ren}}{v_{F}} =
\left(\frac{\partial\mu}{\partial n}\right) \left(\frac{1 + F_0^s}{1+F_1^s}\right).
\end{equation}
Hence, although $(\partial\mu/\partial n)_0$ for the bare system is proportional to the bare particle velocity $v_0$ at the Fermi surface, 
$(\partial\mu/\partial n)_{\rm ext}$ for the extrinsic case is {\em not} proportional to the quasiparticle velocity
because of the presence of the additional Fermi liquid parameter $F_0^s$.

We conclude with a discussion of the possible effects of disorder and correlation on graphene $\partial\mu/\partial n$. 
We believe that correlation, neglected in our Hartree-Fock theory, would introduce only small quantitative
corrections to our calculated results, particularly because of the relatively small values of $r_s$ ($<1$) in graphene.  Thus, our
Fock exchange approximation for graphene $\partial\mu/\partial n$ should quantitatively be an excellent approximation.  Disorder would also introduce
only small quantitative corrections except at low extrinsic carrier densities ($|n| \lesssim 5\times 10^{11}\,{\rm cm}^{-2}$) associated with the so-called
``minimal graphene conductivity" regime, where random charged impurities in the substrate introduce\cite{condmat0610157,condmat0702117} inhomogeneous electron/hole puddles in the
graphene layer which would lead to random spatial variations in $\partial\mu/\partial n$ over $5$ -- $20$ nm ($10$ -- $100$ meV) length (energy) scales.
Finally, finite temperature would have little effect on our results because $E_F > 1000\,{\rm K}$ in the usual density range of experimental interest ($\gtrsim 
5\times 10^{11}\,{\rm cm}^{-2}$).

This work is supported by US-ONR and LPS-NSA.

{\sl Note added:} After submission of this manuscript, we received a preprint\cite{Yacoby_unpubl} reporting an experimental observation of the exchange contribution
to $\partial\mu/\partial n$ that is consistent with our theory. 



\end{document}